\documentclass{aa}
\usepackage{graphics}
\def\ga{\lower.5ex\hbox{$\; \buildrel > \over \sim \;$}}
\def\la{\lower.5ex\hbox{$\; \buildrel < \over \sim \;$}}

\begin{document}      


   \title{$^{12}$CO(1--0) observations of NGC~4848:\\ a Coma galaxy after stripping}

   \author{B.~Vollmer\inst{1,2}, ~J. Braine\inst{3}, C.~Balkowski\inst{2}, 
   V.~Cayatte\inst{2} \and W.J.~Duschl\inst{4,1}}

   \offprints{B.~Vollmer, e-mail: bvollmer@mpifr-bonn.mpg.de}

   \institute{Max-Planck-Institut f\"ur Radioastronomie, Auf dem H\"ugel 69,
   	      D-53121 Bonn, Germany. \and 
              Observatoire de Paris, DAEC,
              UMR 8631, CNRS et Universit\'e Paris 7,
	      F-92195 Meudon Cedex, France. \and
	      Observatoire de Bordeaux, UMR 5804, CNRS/INSU, B.P. 89, 
	      F-33270 Floirac, France. \and
	      Institut f\"ur Theoretische
              Astrophysik der Universit\"at Heidelberg, 
	      Tiergartenstra{\ss}e 15, D-69121 Heidelberg, Germany. 
 	      } 
   	  
   \date{Received 16 February 2001 / Accepted 22 May 2001}

   \authorrunning{Vollmer et al.}
   \titlerunning{$^{12}$CO(1--0) observations of NGC~4848}

\abstract{
We study the molecular gas content and distribution in the Coma cluster spiral galaxy
NGC~4848 which is H{\sc i} deficient and where the remaining atomic gas is 
on one side of the galaxy, presumably because of the strong ram-pressure
in the rich Coma cluster.  Plateau de Bure interferometric CO(1--0) observations
reveal a lopsided H$_2$ distribution with an off-center secondary maximum
coincident with the inner part of the H{\sc i}.  NGC~4848 is not at all 
deficient in molecular gas as it contains 
M$_{\rm H_{2}} \sim 4 \, 10^{9}$~M$_{\odot}$ in the central and inner disk regions.\\
As predicted by earlier calculations, ram-pressure has little influence 
on the dense molecular gas in the inner disk, which appears dynamically normal
at our 2 kpc resolution.  At the interface between the CO and H{\sc i} emission regions, 
about 8~kpc NW of the center, however, strong star formation is present as witnessed
by H$\alpha$ and radio continuum emission.  From the radio and H$\alpha$ fluxes
we estimate the star formation rate in the northern emission region to be close to 
0.3~M$_{\odot}$\,yr$^{-1}$, nearly that of an average quiescent spiral.  
This is the region in which 
earlier Fabry-P\'erot observations revealed a double-peaked H$\alpha$ line,
indicating gas at two different velocities at the same sky position.\\
In order to understand these observations, we present the results of numerical 
simulations of the galaxy-cluster ICM interaction.  We suggest that NGC~4848  
already passed through the center of the cluster about 4\,10$^{8}$ years ago.
At the observed stage ram pressure has no more direct dynamical influence on the
galaxy's ISM. We observe the galaxy when a fraction of the stripped gas is falling 
back onto the galaxy.
Ram pressure is thus a short-lived event with longer-lasting consequences.
The combination of ram-pressure and rotation results in gas at different
velocities colliding where the double-peaked H$\alpha$ line is 
observed. Ram-pressure can thus result, after re-accretion, in displaced molecular 
gas without the H$_2$ itself being pushed efficiently by 
the intracluster medium.  This process, however, requires strong stripping, such that 
only galaxies with radial orbits can be affected as much as NGC~4848.
A scenario where two interactions take place simultaneously is also
consistent with the available data but less probable
on the basis of our numerical simulations.
}

\maketitle

\section{Introduction}

Single dish H{\sc i} 21 cm line observations
show that the spiral galaxies in the central part ($\leq$ 1.5$^{\rm o}$)
of a large galaxy cluster have less atomic gas than field galaxies of the same 
morphological type and optical diameter, i.e. they are H{\sc i} deficient
(see e.g. Giovanelli \& Haynes 1985, Gavazzi 1989).
High resolution observations reveal distortions in their H{\sc i} 
distribution (Bravo-Alfaro et al. 2000).
Despite their reduced atomic gas content, the molecular mass of H{\sc i}
deficient galaxies is comparable to that of corresponding field galaxies 
(Kenney \& Young 1989 for the Virgo cluster; Casoli et al. 1991,
Boselli et al. 1997 for the Coma cluster). 
This suggests that the mechanism which is responsible for the removal
of the atomic gas does not affect the molecular phase. 

The only type of galaxy-cluster interaction which removes the
neutral H{\sc i} gas from the outer disk without modifying the molecular 
gas phase or stellar distribution is ram pressure stripping (Gunn \& 
Gott 1972). Ram pressure is exerted by the hot intracluster medium (ICM)
on the interstellar medium (ISM) of the fast moving galaxy and is
proportional to the galaxy velocity squared and the ICM density.  Spiral
galaxies are recent cluster members (Tully \& Shaya 1984, Solanes 2000). 
Many of them should have eccentric orbits, leading them near the cluster 
center. The gas in clusters is so centrally concentrated that ram-pressure is 
only important close to the center. When the consequences of ram-pressure 
are visible, it means the galaxy has already passed through the center
if it is H{\sc i} deficient ($DEF>0.3$) (Vollmer et al. 2001).  
Combes et al. (1988) and Kenney \& Young (1989) have shown that molecular 
clouds are located too deep in the galaxy's gravitational potential and
have column densities too high to be removed by ram pressure. 
Ram pressure stripping removes only atomic gas clouds with lower 
column densities but it can alter the orbits of the molecular clouds once
much of the H{\sc i} is stripped (Vollmer et al. 2001). 

The effects of
ram pressure depend on the galaxy orbit and the inclination angle $i$
between the orbital and the disk plane. Numerical simulations 
(Vollmer et al. 2001) show that in the case of small $i$
ram pressure not only strips the atomic gas from the outer parts 
of the disk but also pushes a considerable amount of gas
to smaller galactic radii. This leads to a very short-lived enhancement of the
gas surface density in the inner disk which might lead to an enhanced
star formation activity there. Furthermore, gas clouds which are not 
accelerated to the escape velocity during the stripping event
will fall back to the galaxy within a few
10$^{8}$~yr, colliding with the galaxy ISM.  While ram-pressure leads
to a long-term decrease in the gas mass (Kenney \& Young 1989) and star 
formation rate (SFR; Kennicutt 1983), it can temporarily lead to an increase
in the gas density, and thus presumably the SFR, in two ways: ($a$)
for $\sim 10^{7}$~yr during the closest passage to the cluster center or 
($b$) for a few $10^{8}$~yr after the closest passage if re-accretion takes 
place in disk regions where the gas has not been stripped.
We have found the signature of the back-falling gas in one spiral
galaxy in the Virgo cluster (NGC~4522, Vollmer et al. 2000).

The Coma cluster is the densest ($n_{\rm e,\,0} \sim 6\,10^{-2}$~cm$^{-3}$, Hughes
et al. 1989) and most X-ray luminous cluster 
($L_{\rm X} \sim 3\,10^{44}$ erg\,s$^{-1}$, Jones \& Forman 1984) in our neighborhood. 
It has been extensively studied at several wavelengths
(H{\sc i} 21 cm: Giovanelli \& Haynes 1985, Gavazzi 1987;1989, Bravo-Alfaro 
et al. 2000; CO: Casoli et al. 1991; 
optical: Colless \& Dunn 1996, Biviano et al. 1996; H$\alpha$:
Amram et al. 1992; UV: Donas et al. 1995, X-rays: Briel et al. 1992, White
et al. 1993, Vikhlinin et al. 1997) resulting in a detailed view of
this cluster.

In general it is believed that ram pressure stripping quenches star formation
(see e.g. Poggianti 1999). However, Donas et al. (1995) observed an enhancement of the
median UV flux and the fraction of blue star-forming galaxies in a 
ring-like region $\sim$25$'$ ($\sim$0.7 Mpc) from the Coma cluster center. 
The majority of these galaxies are H{\sc i} deficient.
They concluded that a global physical process which leads to the
H{\sc i} deficiency of these cluster galaxies might 
induce star formation in a rich cluster such as Coma.

The evolution of the SFR is a key factor determining the future appearance
of galaxies. Since stars form in molecular clouds, we investigate the effect
of ram-pressure on the molecular gas in the environment where it should be
most clearly discernible -- that of a rich cluster.
We present high-resolution CO(1--0) observations of NGC~4848, an H{\sc i}
deficient UV-bright spiral galaxy in the richest nearby cluster -- Coma.
It is classified as a blue disk galaxy in the 
sample of Bothun \& Dressler (1986). Since it is far away from the cluster
center ($\sim$1~Mpc) and highly H{\sc i} deficient  but still forms stars at
a high rate ($\sim$3~M$_{\odot}$\,yr$^{-1}$), it is an ideal
candidate for the scenario described above where the stripped gas falls back 
onto the galactic disc. We are thus able to test our simulations directly
with observations at multiple wavelengths.

We compare the molecular gas distribution 
and velocity field of NGC~4848 with H{\sc i} and H$\alpha$ interferometric
observations. With the help of our dynamical model simulating 
ram pressure stripping, we investigate the link between the stripping event and
star formation. The structure of this article is the following:
The CO(1--0) observations are presented in Sect.~2--4. They are compared
in Sect.~5 to H band, B band, H{\sc i}, H$\alpha$, and 20~cm continuum images.
In Sect.~6 we discuss the observational data. The numerical model is
presented and compared to the observations in Sect.~7. Alternative scenarios
are discussed in Sect.~8, which is followed by the conclusions (Sect.~9).

We adopt a distance of 100~Mpc for the Coma cluster, corresponding 
to a Hubble constant of about 70~km$\,$s$^{-1}\,$Mpc$^{-1}$.
All velocities are heliocentric ($v=c\,z$).

\section{Observations}

The observations were made with the IRAM interferometer of Plateau de Bure
(see description in Guilloteau et al. 1992). We observed NGC~4848
in April 1999 with the 5D configuration at 112.5674 GHz during 6 hours and in May 1999 
with the 5C configuration during 7 hours.
The antenna half-power primary beam was 45$''$; the primary beam field was
centered on the optical center of NGC~4848. A total bandwidth of 435 MHz 
(1160 km\,s$^{-1}$) centered on v$_{\rm sys}$=7200 km\,s$^{-1}$ was observed 
with a resolution of 2.5 MHz (6.6 km\,s$^{-1}$). 

Data calibration was made in the standard way using the ``CLIC'' software
package (Lucas 1992). The RF passband was calibrated at the start
of each session 
observing 3C273 or 0923+392. The relative phase of the antennas was checked
every 20 mn on the nearby quasars 1308+326 and 1328+307. The rms 
atmospheric phase fluctuations were typically between 10$^{\rm o}$
and 30$^{\rm o}$. We applied a Hanning smoothing to the channels giving
58 channels of width 20 km\,s$^{-1}$. The resulting beam size is
$5.5'' \times 3.6''$. The rms noise of one of these
channels is 2.3 mJy/beam ($\Delta$T$_{\rm B}$=0.012 K).

\section{Results}

\begin{figure}
	\resizebox{\hsize}{!}{\includegraphics{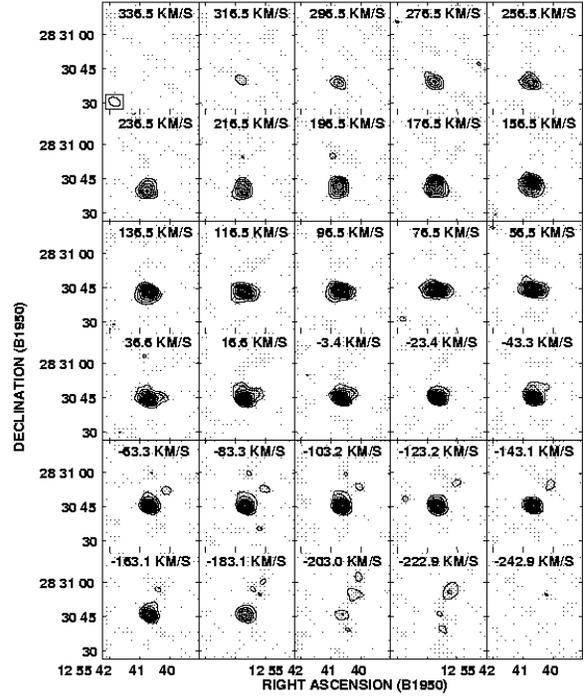}}
	\caption{The $^{12}$CO(1--0) channel maps. The heliocentric velocity 
	relative to v$_{0}$=7200 km\,s$^{-1}$ is indicated in 
	km\,s$^{-1}$ in the upper right of each channel map. The contours
	are 3, 6, 12, 18$\times \sigma$ ($\sigma$=2.3 mJy/beam).  
	} \label{fig:co_10_channels}
\end{figure}

\begin{figure}
	\resizebox{\hsize}{!}{\includegraphics{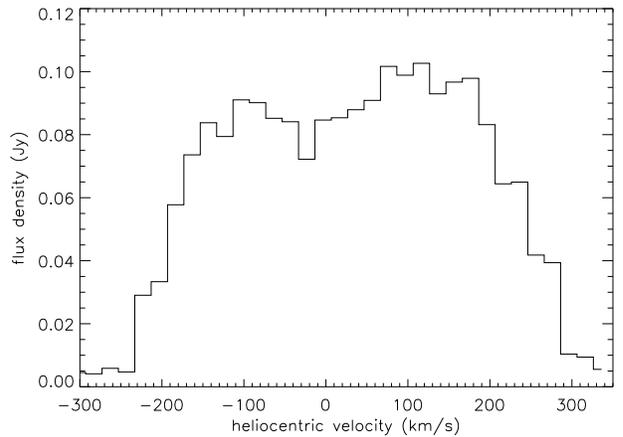}}
	\caption{The integrated spectrum over a box of 20$'' \times 25''$
	covering the whole CO emission. The velocities are centered on 
	v$_{0}$=7200 km\,s$^{-1}$.} \label{fig:co_10_spectrum}
\end{figure}

\begin{figure}
	\resizebox{\hsize}{!}{\includegraphics{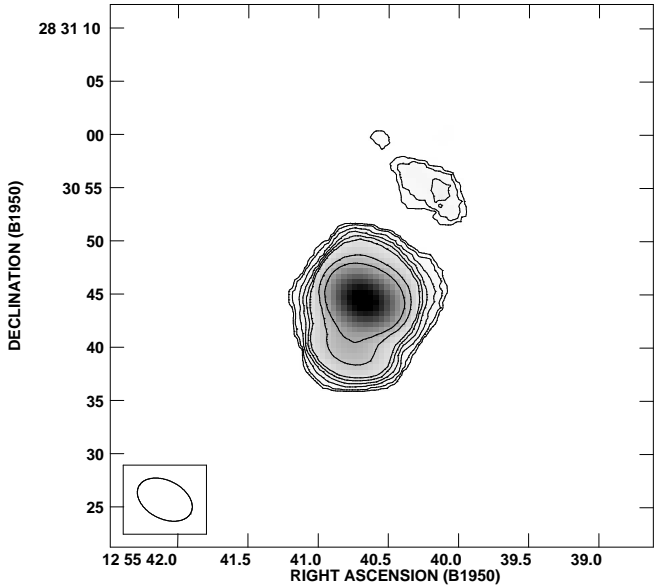}}
	\caption{ $^{12}$CO(1--0) integrated intensity contours of NGC~4848.
	Contours are (1, 2, 4, 6, 8, 10, 20, 30)$\times$ 0.22 Jy/beam$\times$
	km\,s$^{-1}$. The beam is indicated in the lower left corner.
	} \label{fig:co_10_image}
\end{figure}

The physical properties of NGC~4848 can be found in Table~\ref{tab:parameters}.
\begin{table}
      \caption{Physical Parameters of NGC~4848}
         \label{tab:parameters}
      \[
         \begin{array}{ll}
            \hline
            \noalign{\smallskip}
	{\rm Other\ names} & {\rm CGCG~160-055} \\
	 & {\rm UGC~8082} \\
	$$\alpha$$\ (1950)$$^{\rm a}$$ &  12$$^{\rm h}55^{\rm m}40.7^{\rm s}$$\\
	$$\delta$$\ (1950)$$^{\rm a}$$ &  28$$^{\rm o}30'45''$$\\
	{\rm Distance\ D\ (Mpc)} & 100 \\
	{\rm Distance\ to\ cluster\ center\ d\ }($$'$$,\ {\rm Mpc}) & 26,\ 0.75 \\ 
	{\rm Morphological\ type} & {\rm S(B)ab}$$^{\rm a}$$, {\rm Sc}$$^{\rm b}$$ \\
	{\rm Optical\ diameter\ D}_{25}\ ($$'$$)$$^{\rm a}$$ & 1.5\\
	{\rm UV\ magnitude\ mUV},\ \log\,{\rm L}_{UV} ({\rm L}_\odot)$$^{\rm c}$$ & 13.8,\ 10.72\\
	{\rm B}$$_{T}^{0},\ \log\,{\rm L}_{B} ({\rm L}_{\odot\,B})^{\rm a}$$ & 13.52,\ 10.78\\
	{\rm H}$$_{T}^{0}\ ^{\rm d}$$ & 10.78\\
	{\rm IRAS}\ 60$$\mu$${\rm m}\ {\rm (Jy)}$$^{\rm e}$$ & 1.34 \\
	{\rm IRAS}\ 100$$\mu$${\rm m}\ {\rm (Jy)}$$^{\rm e}$$ & 2.60 \\
	\log\ {\rm L(FIR)\ (erg\,s}$$^{-1}$$)$$^{\rm f}$$ & 43.87 \\
	{\rm V}$$_{\rm sys, \, opt, \, hel}$$\ {\rm (km\,s}$$^{-1}$$)$$^{\rm a}$$ & 7227$$\pm$$19\\
	{\rm V_{HI, \,opt}\ (km\,s}$$^{-1}$$)$$^{\rm k}$$ & 7049$$\pm$$10 \\
	\log\ {\rm F(H}$$\alpha$$)\ ({\rm erg}\,{\rm cm}$$^{-2}$$\,{\rm s}$$ ^{-1}$$)$$^{\rm g}$$ & -12.51\\
	\log\ {\rm L(H}$$\alpha$$)\ ({\rm erg\,s}$$ ^{-1}$$)$$^{\rm g}$$ & 41.63\\
	{\rm V}$$_{\rm rot}$$\ {\rm (km\,s}$$^{-1}$$)$$^{\rm h}$$ & 270 \\
	{\rm Position\ angle\ (gas\ kinematics)}$$^{\rm j}$$ & 158$$^{\rm o}$$ \\
	{\rm Position\ angle\ (optical\ image)}$$^{\rm a}$$ & 158$$^{\rm o}$$ \\
	{\rm Inclination\ of\ the\ gaseous\ disk}$$^{\rm h,j}$$ & 75$$^{\rm o} \pm 5^{\rm o}$$ \\
	{\rm S}$$_{\rm CO}$$\ {\rm (Jy\,km\,s}$$^{-1}$$),\ {\rm M}$$_{\rm mol}$$ (M$$_{\odot})^{\rm j}$$ & 41,\ $$4.5 \, 10^9$$ \\
	{\rm S}$$_{\rm HI}$$\ {\rm (Jy\,km\,s}$$^{-1}$$),\ {\rm M}$$_{\rm HI}$$ (M$$_{\odot})^{\rm k}$$& $$0.7,\ 1.7\,10^9$$ \\
	\noalign{\smallskip}
	\hline
	\end{array}
      \]
\begin{list}{}{}
\item[$^{\rm{a}}$] RC3
\item[$^{\rm{b}}$] Dressler (1980)
\item[$^{\rm{c}}$] Donas et al. (1995)
\item[$^{\rm{d}}$] Gavazzi \& Boselli (1996)
\item[$^{\rm{e}}$] Moshir et al. (1990)
\item[$^{\rm{f}}$] Bicay \& Giovanelli (1987)
\item[$^{\rm{g}}$] Gavazzi et al. (1998), $L_{\rm H\alpha}$ is corrected for 1 magnitude extinction
\item[$^{\rm{h}}$] Amram et al. (1992)
\item[$^{\rm{j}}$] this paper
\item[$^{\rm{k}}$] Giovanelli \& Haynes (1985)
\end{list}
\end{table}
The complete $^{12}$CO(1--0) cube is shown in Fig.~\ref{fig:co_10_channels}. 
The integrated spectrum over a box of 20$'' \times 25''$ covering the
whole CO emission can be seen in Fig.~\ref{fig:co_10_spectrum}.
The velocities are centered on v$_{\rm sys}$=7200 km\,s$^{-1}$.
The flux density of the receding part of the galaxy (south) is slightly
higher than that of the approaching side (north).
The total CO flux is $S_{\rm CO}$=41 Jy\,km\,s$^{-1}$. Using the
formula of Kenney \& Young (1989):
\begin{equation}
M_{\rm H_{2}}=1.1\,10^{4} \times D^{2} \times S_{\rm CO}\,
\end{equation}
where $D$ is the distance of the galaxy in Mpc
gives a total mass of molecular gas of $M_{\rm H_{2}}$=4.5\,10$^{9}$
M$_{\odot}$.  This corresponds to a conversion factor of 
$N({\rm H}_2) / I_{\rm CO(1-0)} = 2.8\, 10^{20}$ 
cm$^{-2}\,{\rm K}^{-1}\,{\rm km/s}^{-1}$.
The CO emission distribution calculated using a cutoff of 3$\sigma$ is
shown in Fig.~\ref{fig:co_10_image}.
The major part of the integrated intensity is found in the central part
of the galaxy. The galaxy's rotation is clearly observed in the channel maps.
It ranges from -203 to 316 km\,s$^{-1}$ (with respect to $v_{0}=7200$~km\,s$^{-1}$).
The maximum of the CO emission coincides with the optical galaxy center.
The emission is elongated in the south-east direction corresponding roughly
to the position angle of the galaxy. The most interesting feature
is located in the north-west of the galaxy center. There, an emission region 
which is hardly spatially resolved is detected. The CO flux within this region
is $S_{\rm CO}=3$ Jy\,km\,s$^{-1}$ corresponding to a molecular
hydrogen mass of $M_{\rm H_{2}}=3.3\,10^{8}$ M$_{\odot}$.
This emission region appears in nine different channels
from -243 to -84 km\,s$^{-1}$. In the integrated intensity map its
highest contour is at 0.88 Jy/beam$\times$km\,s$^{-1}$. 

\section{Kinematics}

\begin{figure}
	\resizebox{\hsize}{!}{\includegraphics{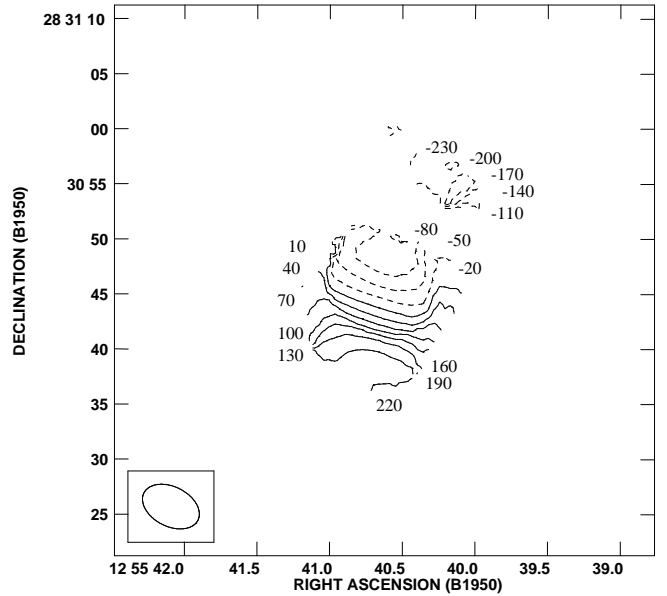}}
	\caption{$^{12}$CO(1--0) velocity field. Contours are
	in steps of 30 km\,s$^{-1}$. The velocities are centered on 
	v$_{0}$=7200 km\,s$^{-1}$. The beam is indicated in the 
	lower left corner. 
	} \label{fig:co_10_velfield}
\end{figure}

\begin{figure}
	\resizebox{\hsize}{!}{\includegraphics{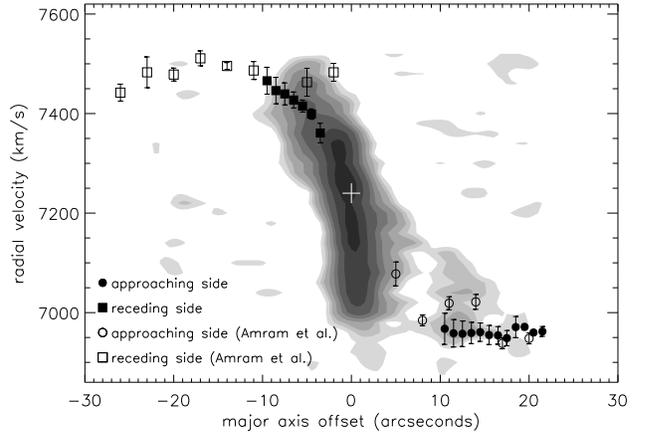}}
	\caption{The p-V diagram along the major axis together
	with the CO and H$\alpha$ rotation curves. 
	The open boxes and circles
	represent the values obtained by Amram et al. (1992). The filled
	boxes and circles represent our $^{12}$CO(1--0) data. 
	The bars indicate the errors on the measurements.
	} \label{fig:co_10_rotcurve}
\end{figure}

The velocity field of the CO emission is plotted in 
Fig.~\ref{fig:co_10_velfield}. 
It is worth noting that within the northern emission region there is a velocity 
gradient of $\Delta v \sim 150$~km\,s$^{-1}$ within only one beam.
This gradient appears to be perpendicular to that of the rotation of the
gas in the disk, but since this is observed within one beamsize it is
not clear if the direction of the gradient is real.
In order to determine the rotation curve we have taken the optical galaxy
center as a first guess for the dynamical center. 
First, we averaged the approaching and receding side of the galaxy 
up to a radius of 30$''$ excluding points within a sector of 
$\pm$30$^{\rm o}$ around the minor axis. The derived dynamical center
coincides with the optical one. The results for the other parameters
are given in Table \ref{tab:parameters}.
The parameters as well as the rotation curves itself are in good 
agreement with the values derived by Amram et al. (1992). 
The corresponding position--velocity diagram together with the our CO rotation 
curve and the H$\alpha$ rotation curve of Amram et al. (1992) are shown in 
Fig.~\ref{fig:co_10_rotcurve}.  
The rotation curve rises steeply in the inner part of the
galaxy. In the outer regions it is flat with a rotation velocity of
$v_{\rm rot}$=270~km\,s$^{-1}$. The rotation curve rises more slowly
at the receding side, which could be due to the deviation of the galaxy's
dynamical major axis from the galaxy's position angle and/or
due to beam smearing.

\section{Comparison with other wavelengths}

\subsection{Near Infrared (H band)}

Gavazzi et al. (1996) observed NGC~4848 in the H band (1.65~$\mu$m) 
with a resolution of $\sim 2''$.
\begin{figure}
	\resizebox{\hsize}{!}{\includegraphics{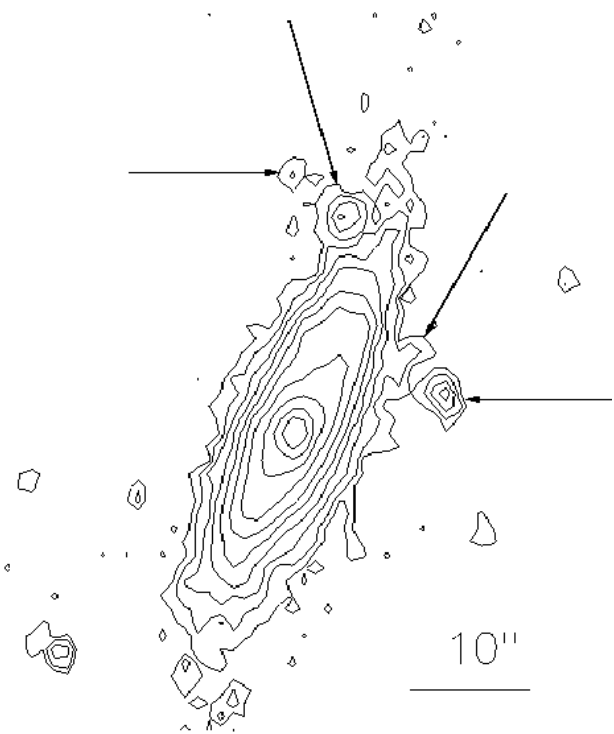}}
	\caption{H band image of NGC~4848 with a resolution of $\sim 2''$.
	The contours are from 17 to 22~mag\,arcsec$^{-2}$ in steps of 
	0.5~mag\,arcsec$^{-2}$. The arrows indicate four distinct spherical 
	emission regions.
	} \label{fig:Hband}
\end{figure}
Fig.~\ref{fig:Hband} shows an almost unperturbed large scale disk, i.e.
the isophots are regular and do not show a major perturbation. Since the
H band traces the old stellar population, we can conclude that the galaxy 
has not undergone a recent major gravitational interaction. However,
its gas distribution and kinematics traced by the CO, H{\sc i}, and H$\alpha$
emission distributions and velocity fields show major asymmetries and perturbations
in the north. This suggests that an ICM--ISM interaction has selectively
disturbed the ISM of NGC~4848. Nevertheless, there could be a minor
perturbation of the old stellar population in the northern
part of the galaxy where an alignment of four distinct spherical emission regions, 
which runs at an angle of $\sim 30^{\rm o}$ across the main stellar disk,
can be distinguished (arrows in Fig.~\ref{fig:Hband}). 
This feature will be discussed in Section~\ref{sec:tdwarf}.

\subsection{Optical B band image}

\begin{figure}
	\resizebox{\hsize}{!}{\includegraphics{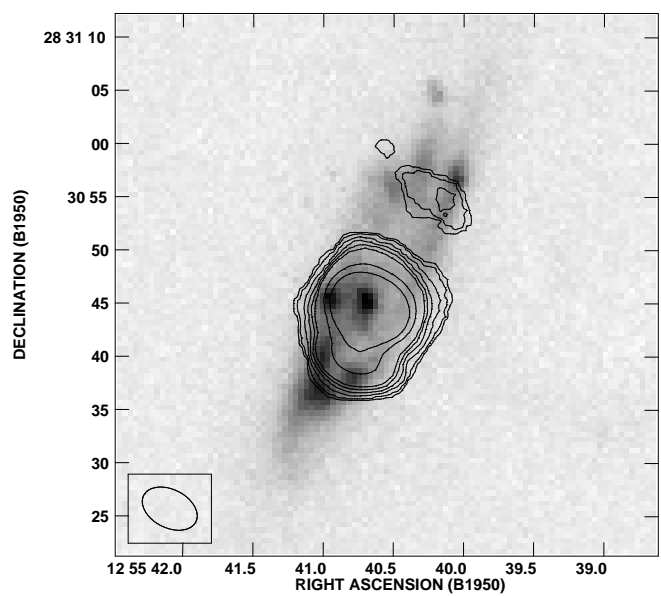}}
	\caption{Greyscales: B band CCD image of NGC~4848 with a resolution of
		0.5$''$/pixel. Contours: CO emission, for contour levels see
		Fig.~\ref{fig:co_10_image}. The CO beam is indicated
		in the lower left corner.
	} \label{fig:co_10_bband}
\end{figure}

Gavazzi et al. (1990) obtained a CCD image of NGC~4848
with a resolution of 0.5$''$ in the B band.
This image is shown in Fig.~\ref{fig:co_10_bband} together with our CO
map.
The B band image shows a blue galaxy center and a ring-like structure 
which is not centered on the galaxy center but shifted $\sim$7$''$ to the north.
There are four bright blue spots in the
southern part of this ring and one bright blue spot in the north.
The southern part of the ring is located at the outer edge of the
southern elongation observed in CO. The bright blue spot in the north
lies at the northern edge of the detached CO emission region. 
Moreover, there are two faint blue emission regions located at the periphery of 
the CO emission region and another one far away to the north.

\subsection{H{\sc i}}

\begin{figure}
	\resizebox{\hsize}{!}{\includegraphics{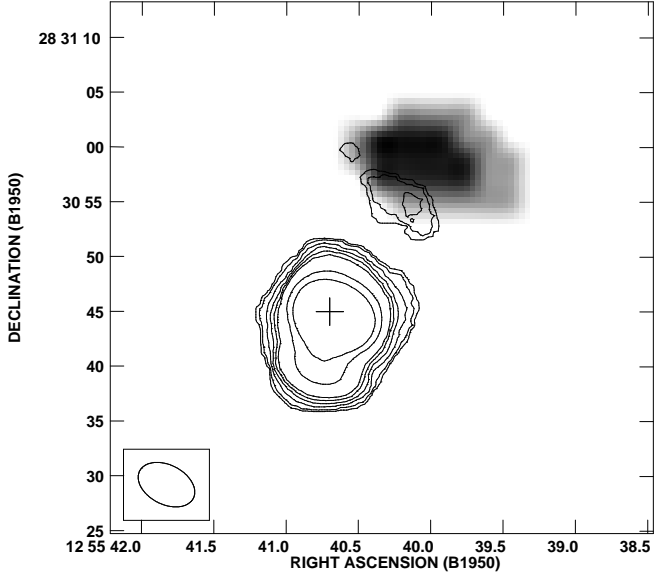}}
	\caption{Greyscales: H{\sc i} distribution of 19$''$ resolution.
		Contours: CO emission, for contour levels see
		Fig.~\ref{fig:co_10_image}. The CO beam is indicated
		in the lower left corner.
	} \label{fig:co_10_HI}
\end{figure}

\begin{figure}
	\resizebox{\hsize}{!}{\includegraphics{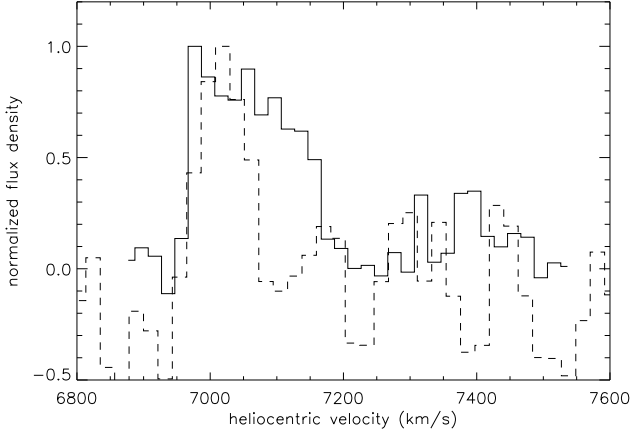}}
	\caption{Solid line: normalized CO integrated spectrum
		of the northern emission region.
		Dashed line: normalized H{\sc i} integrated spectrum of the 
		northern emission region. 
	} \label{fig:co_10_HI_spectra}
\end{figure}

Bravo-Alfaro et al. (1999, 2000) observed several fields in the Coma
cluster in the H{\sc i} 21 cm line with the VLA C array.
We calibrated and treated the field containing
NGC~4848 for a second time in order to obtain the
maximum resolution. The continuum was removed using all velocity channels 
between 6800 km\,s$^{-1}$ and 7800 km\,s$^{-1}$.
We obtained a final rms noise in one 20 km\,s$^{-1}$ wide channel
of 0.3 mJy per 19$''$ beam. Fig.~\ref{fig:co_10_HI} shows the H{\sc i}
distribution as greyscales and the CO emission as contours.
As already observed by Bravo-Alfaro et al. (2000) the H{\sc i} 
emission is located in the north. With an H{\sc i} beamsize of
$\sim 20''$ it is not spatially resolved. The H{\sc i} emission region is  
located at a larger distance from the galaxy center than the CO emission region. 
Since this emission region appears in 5 different velocity channels 
above the 2$\sigma$ level, we are confident
that this detection is real. It represents the neutral gas with the highest
column density. The lower resolution map obtained by Bravo-Alfaro et al.
(2000) shows that the emission extends northwards.
In Fig.~\ref{fig:co_10_HI_spectra}
we compare the spectra of the H{\sc i} and CO emission regions integrated over
the area where emission is detected. 
Both spectra show emission in the same velocity range between 7000
and 7100 km\,s$^{-1}$, the CO line being two times larger.
This confirms that both detections are real. Furthermore, the
integrated H{\sc i} spectrum of Giovanelli \& Haynes (1985) has the same
shape and width as our CO spectrum. Thus, the northern CO 
emission region and the H{\sc i} emission region are very probably
associated. Given the small errors on the optical and H{\sc i} velocities
($\sim$20~km\,s$^{-1}$) compared to their difference 
($\Delta v \sim 180$~km\,s$^{-1}$, see Tab.~\ref{tab:parameters}),
we think that the offset between the optical and H{\sc i} velocities is real.

The H{\sc i} flux in the C-array VLA map of this region is
$S_{\rm HI}$=0.05 Jy\,km\,s$^{-1}$ (M$_{\rm HI}$=1.2\,10$^{8}$ M$_{\odot}$),
whereas the single-dish Arecibo observations show that much more low surface 
density H{\sc i} is present (M$_{\rm HI}$=1.7\,10$^{9}$ M$_{\odot}$).  
The CO flux of the northern emission region is 
$S_{\rm CO}$=3 Jy\,km\,s$^{-1}$ (M$_{\rm H_{2}}$=3.3\,10$^{8}$ M$_{\odot}$),
which represents $\sim$7\% of the total CO flux.
More sensitive H{\sc i} observations are necessary to determine whether the
gas is chiefly atomic or molecular in this region.

\subsection{H$\alpha$}

\begin{figure}
	\resizebox{\hsize}{!}{\includegraphics{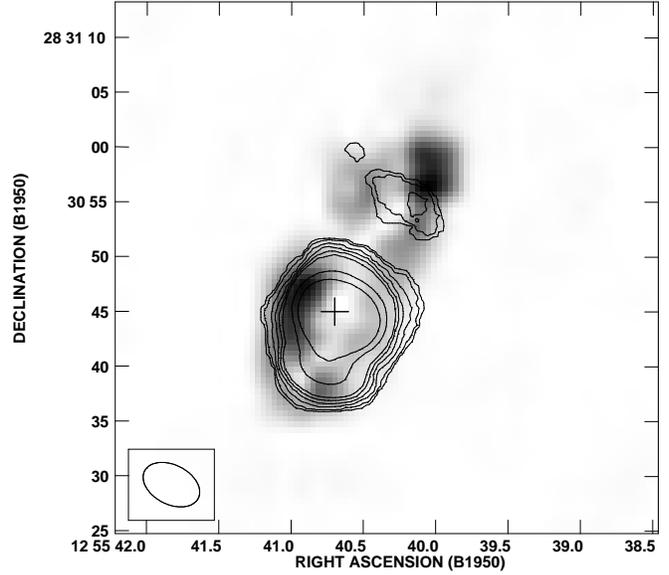}}
	\caption{Greyscales: H$\alpha$ distribution with a resolution of 2$''$.
		Contours: CO emission, for contour levels see
		Fig.~\ref{fig:co_10_image}. The CO beam is indicated
		in the lower left corner.
	} \label{fig:co_10_Ha}
\end{figure}

We compare here our CO data with the H$\alpha$ Fabry-P\'erot
observations of Amram et al. (1992).
Their map is shown as greyscales together with the CO data as contours
in Fig.~\ref{fig:co_10_Ha}.
We determined the coordinates of the galaxy center of the H$\alpha$ image 
by fitting it to the B band image.
The H$\alpha$ emission distribution shows a ring structure very similar
to the one observed in the B band image. However, it contains only two bright
spots in the south west and in the north where a lot of massive stars
are formed. The other three bright spots in the south of the B band image
do not have counterparts in H$\alpha$. The central hole is probably
due to a too severe continuum subtraction in this region as a comparison
with the H$\alpha$ image of Gavazzi et al. (1998) indicates.
The northern bright spot
is located near the CO emission region. The uncertainty in the determination
of the galaxy center coordinates in the H$\alpha$ image
does not allow us to conclude if both emission regions coincide exactly.
We conclude here that both are located very close to each other
within the same region. The northern H$\alpha$ emission region
contains $\sim$20\% of the total H$\alpha$ emission.  In the south,
the outer edge of the CO elongation nicely fits the H$\alpha$ distribution.

\subsection{H$\alpha$/H$\alpha$ continuum}

\begin{figure}
	\resizebox{\hsize}{!}{\includegraphics{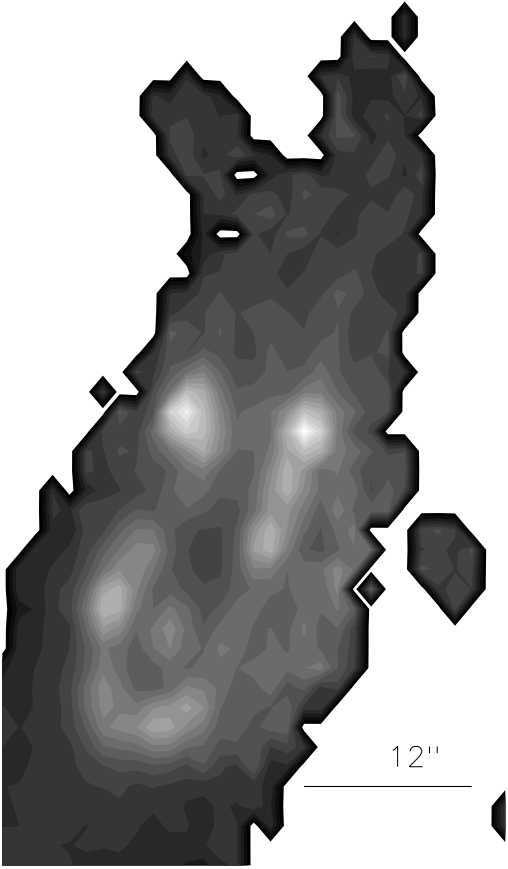}}
	\caption{Ratio between the H$\alpha$ line emission and its continuum.
	The contours are from 0.2 to 4 in steps of 0.2. We have
	cut the south--eastern part of the galaxy because of straylight 
	problems due to a nearby star (Gavazzi et al. 1998).
	} \label{fig:Ha_cont}
\end{figure}

Gavazzi et al. (1998) obtained a narrow band H$\alpha$ CCD image of NGC~4848
with a resolution of 1.2$''$. We show in Fig.~\ref{fig:Ha_cont} the ratio between
the H$\alpha$ line emission and its continuum, in order to study the nature
of the off--plane NIR emission regions (Fig.~\ref{fig:Hband}).
The image shows clearly the ring of highest H$\alpha$ emission 
(see Fig.~\ref{fig:co_10_Ha}). Moreover, the western part of the galaxy next to the
H$\alpha$ ring has an enhanced H$\alpha$/continuum ratio. The NIR emission regions, 
which are running at an angle of $\sim 30^{\rm o}$ across the galaxy's disk 
(see Fig.~\ref{fig:Hband}), have a low H$\alpha$/continuum ratio, similar to that
of the extended disk. The north western H$\alpha$ emission maximum is not
aligned with the NIR emission regions but $2.5'' \simeq 1$~kpc offset to the south.
On the contrary, in the scenario of a tidally stretched dwarf whose ISM is 
interacting with the ISM of NGC~4848, one would expect (i) an alignment of
all these regions and (ii) a higher H$\alpha$/continuum ratio in one of the two 
brightest NIR off--plane regions (see, e.g., Clemens et al. 2000).

\subsection{20 cm}

\begin{figure}
	\resizebox{\hsize}{!}{\includegraphics{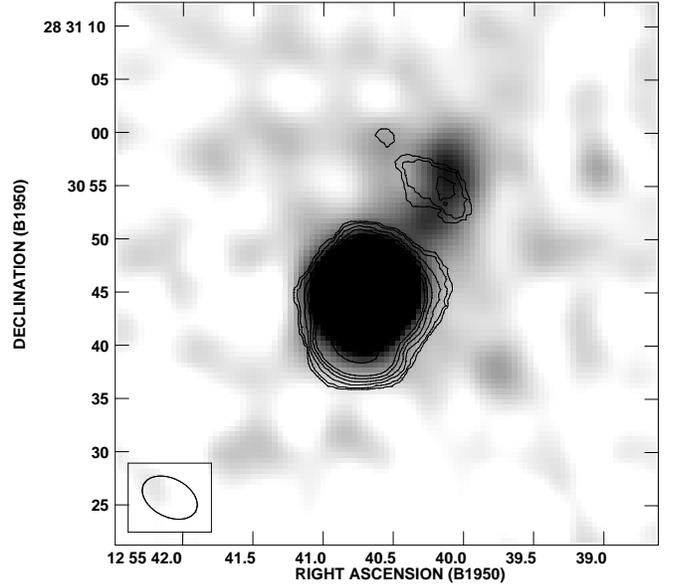}}
	\caption{Greyscales: 20 cm continuum distribution of 5$''$ 
		resolution. Contours: CO emission, for contour levels see
		Fig.~\ref{fig:co_10_image}. The CO beam is indicated
		in the lower left corner.
	} \label{fig:co_10_20cm}
\end{figure}

The Very Large Array (VLA) carries out a systematic survey of the northern 
sky at 20 cm wavelength in the B configuration 
(FIRST; Becker et al. 1994). These maps have 1.8$''$ 
pixels, a typical rms of 0.15 mJy, and a resolution of 5$''$. 
The 20 cm continuum 
map of NGC~4848 together with the CO emission map is shown in 
Fig.~\ref{fig:co_10_20cm}.
The 20 cm continuum map shows enhanced emission to the north west. It 
coincides exactly with the maximum of the CO emission in the northern
emission region. The southern elongation of the CO emission does not have a
counterpart in the 20 cm continuum map. 

\section{Discussion of the observations}

Many of the physical properties of NGC~4848 are comparable to those 
of Virgo cluster Sab galaxies with the same blue luminosity and the
same H{\sc i} deficiency (Kenney \& Young 1989):
\begin{itemize}
\item
S(60 $\mu$m)/S(100$\mu$m)=0.52
\item
ratio of molecular hydrogen mass to optical area:
log($M_{\rm H_{2}}/D_{\rm opt}^{2}$)=0.37~M$_{\odot}\,$pc$^{-2}$
\item
ratio of total gas mass to optical area:
log\big($(M_{\rm HI}+M_{\rm H_{2}})/D_{\rm opt}^{2}$\big)=0.51~M$_{\odot}\,$pc$^{-2}$
\item
ratio of H{\sc i} to H$_{2}$ flux:
log($S_{\rm CO}/S_{\rm HI}$)=1.8\ ,
\end{itemize}
whereas its H$\alpha$ line flux is a factor of at least two higher 
than that of a Virgo cluster spiral of the same morphological type, 
the same diameter, and the same H{\sc i} deficiency. 
In addition, its equivalent width is about 3 times higher than that
of ``normal'' field Sab galaxies (Kennicutt 1983) and resembles
that of an Sc spiral with a star formation rate of 
\begin{equation}
{\rm SFR}\sim \frac{L({\rm H}_{\alpha})}{1.26\,10^{41}
{\rm erg\,s}^{-1}} \sim 4.5\,10^{-44}\ {\rm L(FIR)} \sim 
3\ {\rm M}_{\odot}\,{\rm yr}^{-1}
\end{equation}
(Kennicutt et al. 1994, Kennicutt 1998).

Calculating the CO deficiency
\begin{equation}
{\rm Def}_{\rm CO} = {\rm log}\,M(H_{2})_{\rm e} - {\rm log}\,M(H_{2})_{\rm o}\ ,
\end{equation}
where $M(H_{2})_{\rm o}$ is the observed and $M(H_{2})_{\rm e}$
the expected molecular mass (see Boselli et al. 1997) gives
negative values, {\it i.e.} it is not CO deficient.

NGC~4848 is a H{\sc i} deficient galaxy (Def$_{\rm HI}\ga 0.5$; 
Gavazzi 1989; Giovanelli \& Haynes 1985), i.e. it has lost about 2/3 
of its atomic gas. The fact that the disk to bulge ratio in the H band appears
to be small (Fig.~\ref{fig:Hband}) also indicates that NGC~4848
was likely an Sc galaxy before entering the cluster.
Since there is no identified  major optical companion neither
a major asymmetry in the H band image, a major
gravitational interaction can be excluded.
The alignment of spherical emission regions in the NIR and all visible
wavelengths might represent a tidally sheared dwarf galaxy which crosses the disk 
of NGC~4848. If this dwarf galaxy has been
gas rich this could have an influence on the gas distribution of NGC~4848.
But since the dwarf galaxy is very small compared to NGC~4848 (Fig.~\ref{fig:Hband}),
it is highly improbable that it is responsible for the displacement of more than 
10$^{9}$~M$_{\odot}$ of atomic gas to the north of NGC~4848. Therefore, ram pressure
stripping is the most probable cause for the atomic gas removal. We will 
discuss the scenario of a tidally stretched dwarf further in Sect.~\ref{sec:tdwarf}.

Numerical simulations (Vollmer et al. 2001) indicate 
that H{\sc i} deficient galaxies (Def$_{\rm HI} > 0.3$) 
which are strongly affected by ram pressure stripping and which are located
far away from the cluster center are observable most probably after their closest
passage to the cluster center. Ram pressure $p_{\rm ram}=\rho_{\rm ICM}
v_{\rm gal}^{2}$ is only high enough near the cluster center, 
where the ICM density $\rho_{\rm ICM}$ and the galaxy's velocity $v_{\rm gal}$
increase rapidly, to overcome the restoring gravitational forces.
Thus, ram pressure only acts in the immediate vicinity of the cluster center.
In addition, it needs only several 10$^{7}$~yr to accelerate the 
H{\sc i} clouds to their escape velocity (Murakami \& Babul 1999), 
thus stripping happens very rapidly. 
Since a galaxy passes the cluster center
with a very high velocity ($v_{\rm gal}>2000$~km\,s$^{-1}$), the
probability of observing it when stripping sets in
is very low. It is therefore most likely that we observe an H{\sc i}
deficient galaxy, which has a distorted H{\sc i} distribution, when
it comes out of the cluster center. We will adopt this view in this work.
Furthermore, Vollmer et al. (2001) showed that there is re-accretion of
stripped material, which has not been accelerated to the escape
velocity, between 2 and 5\,10$^{8}$ yr after the closest
passage. This infalling gas hits the galactic disk leading to frequent
cloud--cloud collisions. The compression of the ISM due to these
collisions might lead to an enhancement of the local star formation 
activity where the accreted gas falls on the disk.

NGC~4848 is located at 26$'$($\sim$0.75 Mpc) from the cluster center.
Assuming a mean velocity of 1800 km\,s$^{-1}$, the closest passage was
about 4\,10$^{8}$ yr ago. Therefore we assume that the galaxy is in the stage
where material is falling back on its disk.
The shocks produced by the encounters of infalling clouds with the ISM
will compress the gas and star formation will take place. 
Given this picture, we can interpret the observational data shown above.

The H{\sc i} and CO maps show emission in the north of the galaxy. This means
that the galaxy's gas content of highest column density 
in the disk is located in the north of the galaxy center. 
We propose that the compression leads to a phase transition from 
atomic to molecular gas in the region where the infalling gas
hits the disk, i.e. in the north.
 
\begin{figure}
	\resizebox{\hsize}{!}{\includegraphics{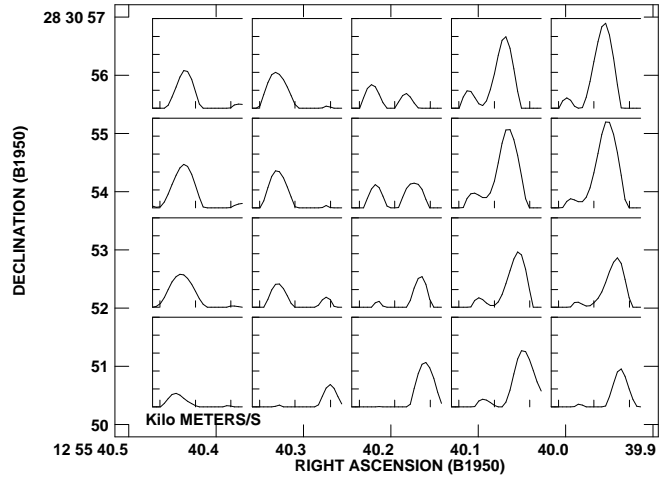}}
	\caption{Spectra of the H$\alpha$ emission region in the north
	of the galaxy; $x$-axis: velocity (6940 to 7468~km\,s$^{-1}$); 
	$y$-axis: H$\alpha$ emission intensity in arbitrary units
	(data cube from Amram et al. 1992). 
	} \label{fig:counterrotation}
\end{figure}

The shift of the H$\alpha$ emission ring to the north indicates an enhanced 
massive star formation in the north of the galaxy. 
However, the extended old disk component is symmetric (Gavazzi et al. 1990).
The bulk of the neutral gas moves around the galaxy center with an almost constant
rotation curve while the ionized gas content shows an important deviation:
in the north western part of the disk the H$\alpha$ emission has a double 
peaked profile (Fig.~\ref{fig:counterrotation} extracted from the H$\alpha$ data 
cube of Amram et al. 1992).
The peak at lower velocities (left) is due to the regular
rotation of the gas within the galaxy following the rotation curve
(Fig.~\ref{fig:co_10_rotcurve}). The stronger peak (right) at higher 
velocities is separated by $\sim$+100 km\,s$^{-1}$ from the first one,
which corresponds to the disk rotation. 
The H$\alpha$ line flux in this region represents 
$\sim20\%$ of the total H$\alpha$ line flux. 

We interpret
this feature as the consequence of infalling gas  
which collides with the ISM within the galaxy. The cloud--cloud collisions 
lead to the compression of the gas and the magnetic field. 
This gives rise to an enhanced star formation observed in the H$\alpha$ map.
The enhanced star formation activity together with the compressed
magnetic field leads to the northern enhancement in the 20~cm continuum map.
The enhanced number of young and hot O/B stars enhanced the ambient 
UV radiation field in the norther part of the galaxy, leading
to an increase of the gas temperature. This effect, together with the
increase of the molecular gas mass due to shock compression during the
re-accretion, increases the $^{12}$CO(1--0) emissivity. The line emission 
thus becomes observable in the northern part of the outer disk.

In the next Section we present a detailed dynamical simulation of the
infalling galaxy including ram pressure stripping in order
to test the hypothesis made above.

\section{Dynamics}

\subsection{The numerical model\label{sec:model}}

Very few simulations have been done to quantify ram pressure stripping 
using Eulerian hydrodynamic (Takeda, Nulsen, \& Fabian 1984, 
Gaetz, Salpeter \& Shaviv 1987; Balsara, 
Livio, \& O'Dea 1994) or SPH codes (Tosa 1994; Abadi, Moore, \& Bower 1999).
All methods have their advantages and their limits.
Since we need to model an extreme density gradient between the ICM and the ISM 
in three dimensions, one would need a very high number of SPH particles
or a very high resolution for Eulerian grid codes which makes
calculations very memory, and time, consuming. The sticky particles code
we use has the advantage that the effects of the ICM are directly
implemented as an additional acceleration of the clouds which are exposed
to ram pressure. The gas viscosity is due to direct (collisions) or indirect
(gravitational) interactions between the clouds. We are the first to
apply a time-dependent ram pressure profile, i.e. a Gaussian profile.
The consequences of the decrease of ram pressure after the galaxy's closest
passage to the cluster center is that material which has not been accelerated
to the escape velocity falls back to the galaxy. This has not been observed
in simulations before, because of the use of a constant, time-independent ram 
pressure. 

\begin{figure}
	\resizebox{\hsize}{!}{\includegraphics{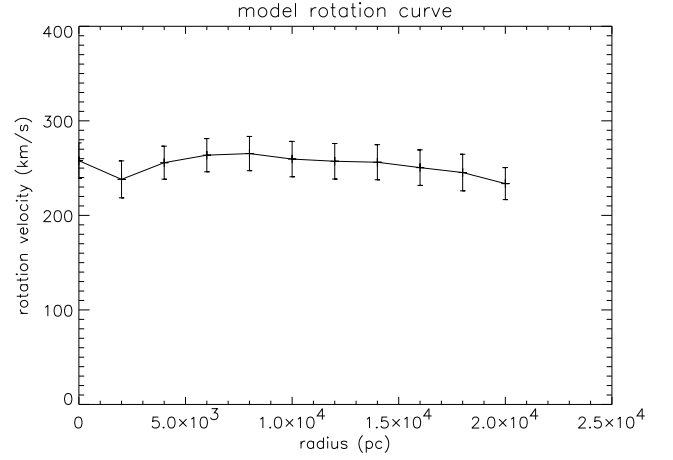}}
	\caption{The model rotation curve. The error bars
	indicate the velocity dispersion of the particles.}
	\label{fig:rotcurve}
\end{figure}

We used the three-dimensional N-body code described in detail in Vollmer
et al. (2001). The particles represent gas cloud complexes which are 
evolving in an analytically given gravitational potential of the galaxy.
This potential consists of two spherical parts: the dark matter halo
and the stellar bulge, and a disk potential (Allen \& Santill\'an 1990). 
The model parameters are: (i) halo: $a_{3}$=12~kpc, 
$M_{3}=2.1\,10^{11}$~M$_{\odot}$, (ii) bulge: $b_{1}$=400~pc, 
$M_{1}=2.8\,10^{10}$~M$_{\odot}$, (iii) disk: $a_{2}$=5~kpc (disk scale
length), $b_{2}$=250~pc (height from the disk plane), 
$M_{2}=1.7\,10^{11}$~M$_{\odot}$. 
The outcoming velocity field has a constant rotation 
curve of $v_{\rm rot} \sim$250 km\,s$^{-1}$ (Fig.~\ref{fig:rotcurve}).
The optical diameter of NGC~4848 $D\sim 44$~kpc corresponds to 
$\sim$9 disk scale lengths.
10\,000 particles of different masses are rotating within this gravitational 
potential. The cloud mass distribution is $n(m)\propto m^{-1.5}$ (Knude 1981)
within a range of $\sim 10^{4}$ to $5\,10^{6}$~M$_{\odot}$.

A radius which depends on the mass is attributed to each particle. 
During the disk evolution they can have inelastic collisions. 
The outcome of these collisions is simplified following Wiegel (1994):
\begin{itemize}
\item
for $r_{1}-r_{2} < b < r_{1}+r_{2}$:\\ fragmentation
\item
for $b \le r_{1}-r_{2}$ and $v_{\rm esc} > v_{\rm f}$:\\ 
mass exchange
\item
for $b \le r_{1}-r_{2}$ and $v_{\rm esc} \le v_{\rm f}$:
\\ coalescence,
\end{itemize}
where $b$ is the impact parameter, $r_{1}$ and $r_{2}$ the cloud radii, 
$v_{\rm f}$ the final velocity difference, and $v_{\rm esc}$ is the escape
velocity. This results in an effective disk viscosity. 
The next neighbor search and selfgravitation between the clouds
is done with a three-dimensional treecode (Barnes \& Hut 1986).

As the galaxy moves through the ICM its clouds are accelerated by
ram pressure. Within the galaxy's inertial system its clouds
are exposed to a wind coming from the opposite direction of the galaxy's 
motion through the ICM. 
The effect of ram pressure on the clouds is simulated by an additional
force on the clouds in the wind direction. Only clouds which
are not protected by other clouds against the wind are affected.

When the galaxy approaches the cluster center, its velocity increases.
At the same time the surrounding ICM density increases. This leads
to an increase of the ram pressure on the ISM clouds $p_{\rm ram}=
\rho_{\rm ICM} v_{\rm gal}^{2}$, where $\rho_{\rm ICM}$ is the ICM
density and $v_{\rm gal}$ is the velocity of the galaxy.
We take this evolution of $p_{\rm ram}$ into account in adopting
the following profile $p_{\rm ram}(t)=28 \rho_{0} v_{0}^{2} {\rm exp}
\big(-(t/1.2\,10^{8})^{2}\big)$, where $\rho_{0}=10^{-4}$ cm$^{-3}$ and
$v_{0}$=1000 km\,s$^{-1}$, where $t=0$~yr is the time of the galaxy's closest
approach to the cluster center.

With the scenario in mind we have taken a snapshot of the evolved galaxy
400~Myr after the closest passage to the cluster center.
Vollmer et al. (2001) have shown that the resulting H{\sc i} deficiency 
increases with increasing inclination angle $i$ between the galaxy's disk 
and its orbital plane, whereas the fraction of re-accreted material increases with
decreasing $i$. In our case, a considerable 
fraction of stripped gas clouds are accelerated by ram pressure to velocities
below the escape velocity.
In addition, for small $i$ the re-accreted material hits 
the disk mainly at one single point. For increasing $i$ the area 
where re-accretion takes place increases and the gas falls
from the $z$ direction onto the disk. For the comparison
with observations we have therefore chosen $i$=20$^{\rm o}$ and a ram pressure
maximum of 28$\rho_{0}\,v_{0}^{2}$. This satisfies all necessary conditions: 
a high stripping efficiency, a relatively high re-accreted mass fraction 
and a small area where the re-accreted cloud complexes hit the disk ISM.
The final H{\sc i} deficiency of this simulation is Def$_{\rm HI}$=0.7.

\begin{figure}
	\resizebox{\hsize}{!}{\includegraphics{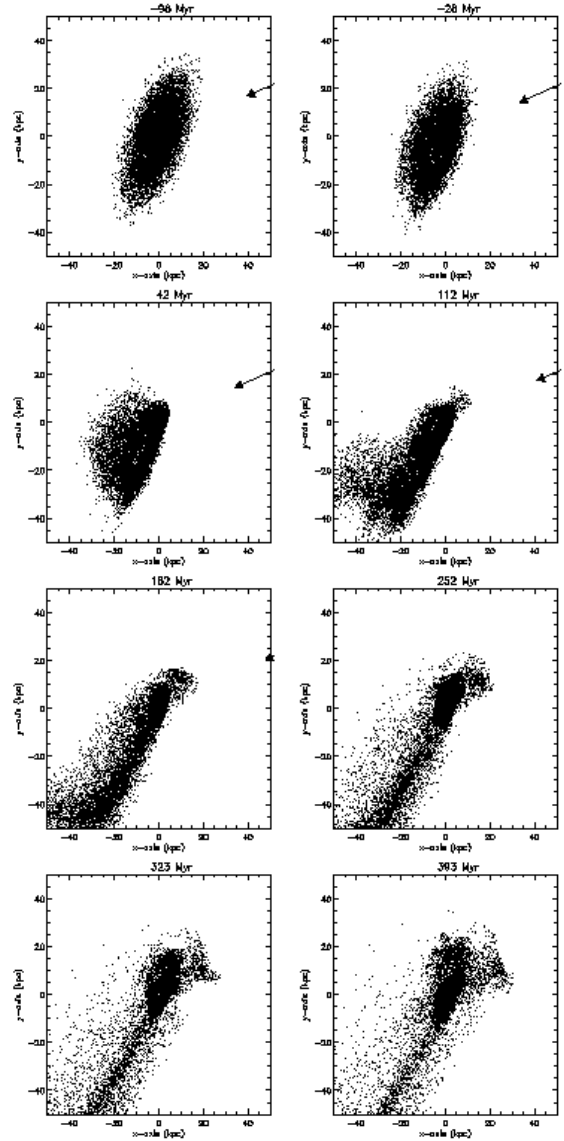}}
	\caption{Snapshots of the galaxy evolution. 
	The elapsed time is indicated at the top of each panel.
	The position and inclination angle of NGC~4848 are used.
	The galaxy rotates clockwise. It is moving in the north-west
	direction, i.e. the wind is coming from the north-west
	(indicated by the arrows). The length of the arrow is proportional 
	to ram pressure ($\rho v^{2}$).}
	\label{fig:simulation1}
\end{figure}

Fig.~\ref{fig:simulation1} shows the evolution of the model galaxy in
timesteps of $\Delta t \sim 70$~Myr.
The ram pressure maximum occurs at $t$=0~yr.

Were NGC~4848 falling into the cluster core, it would
be even less disturbed than in the first panel of Fig.~\ref{fig:simulation1},
which is for 98~Myr before reaching the core.
This is consistent with the conclusion of Vollmer et al. (2001) that highly
distorted, H{\sc i} deficient galaxies which are most likely located far away
from the cluster center (and in particular NGC~4848) are
coming out of the Coma cluster core.

At $t \sim 100$~Myr before the galaxy's closest passage to
the cluster center a density enhancement grows in the direction of
the galaxy's motion. The gas of the outer H{\sc i} disk is driven
away from the galaxy. Since it still has the angular momentum of
rotation, it forms a ring-like structure. At the closest passage
to the cluster center an accelerated arm in the north-east and
a decelerated arm in the south are formed. Both arms rotate 
counter-clockwise. The rotation of the southern arm is stopped by
the fading ram pressure. At $t \sim 200$~Myr after the
galaxy's closest passage to the cluster center the stripped gas
which has been accelerated to velocities remaining below the escape velocity
begins to fall back onto the galaxy following both arms. The infalling
gas flows cross in the north-west of the galaxy where they form
an expanding gas shell. 

\begin{figure*}
	\resizebox{\hsize}{!}{\includegraphics{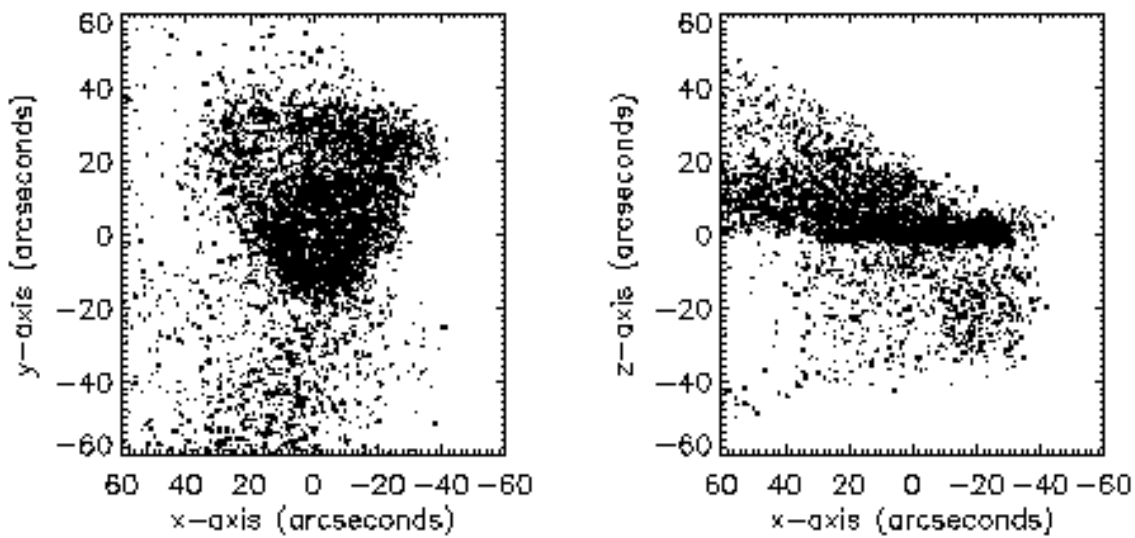}}
	\caption{Model gas distribution.
	Left: Snapshot of the simulation 
	at $t=400$~Myr after the closest passage of the cluster center.
	The sizes of the circles correspond to the sizes of the cloud 
	complexes. The clouds are projected on the $x$-$y$-plane.
	Right: The clouds are projected on the $x$-$y$-plane.
	} \label{fig:model_distribution}
\end{figure*}

The model snapshot at $t=400$~Myr after the closest passage of the 
cluster center is shown in Fig.~\ref{fig:model_distribution}.
The direction of the galaxy's motion indicated in Fig.~\ref{fig:simulation1}.
The hole in the center of the galaxy is artificial, because the clouds
in the very center are arbitrarily removed from the model. Since the
timestep is roughly inversely proportional to the acceleration 
$a$, $\Delta t \propto a^{-1} \sim R\,v_{\rm rot}^{-2}$, 
where $R$ is the galactic distance and we assume 
$v_{\rm rot}$=const, this procedure 
insures a large enough timestep ($\Delta t \sim 10^{4}$~yr). 
The galaxy's main H{\sc i} disk 
is truncated at a diameter of $D_{\rm HI} \sim 15$ kpc.
Assuming an optical diameter of 40~kpc for the model galaxy, 
this gives a ratio between the H{\sc i} and the optical diameter
of  $D_{\rm HI}/D_{\rm opt} \sim 0.4$, which fits in the
corresponding plot of Cayatte et al. (1994).
If we assume the same stripping mechanism acting in the Coma cluster
as for the Virgo cluster, we expect a ratio between the H{\sc i} and the
optical diameter to be $D_{\rm HI}/D_{\rm opt} \sim 0.4-0.5$.
With the optical diameter of NGC~4848 this leads to a predicted H{\sc i}
diameter of $D_{\rm HI}^{\rm predicted}  \sim 20$ kpc (40$''$),
which is only slightly larger than the model H{\sc i} diameter.
The detected H{\sc i} emission region is located at a distance from the
galaxy center of $\sim 15''$, thus well within this predicted diameter.

A shell of high surface density H{\sc i} gas is located in 
the north (Fig.~\ref{fig:model_distribution}, left panel).
It spans an angle of $\sim 45^{\rm o}$. This shell is located mainly
above the disk plane (Fig.~\ref{fig:model_distribution}, right panel).
It consists of gas which has been removed from the disk by ram pressure 
in the wind direction (20$^{\rm o}$ off plane). This
material is now falling back into the galaxy. In addition, a tenuous gas
component can be found north and south of the galaxy.

\begin{figure*}
	\resizebox{\hsize}{!}{\includegraphics{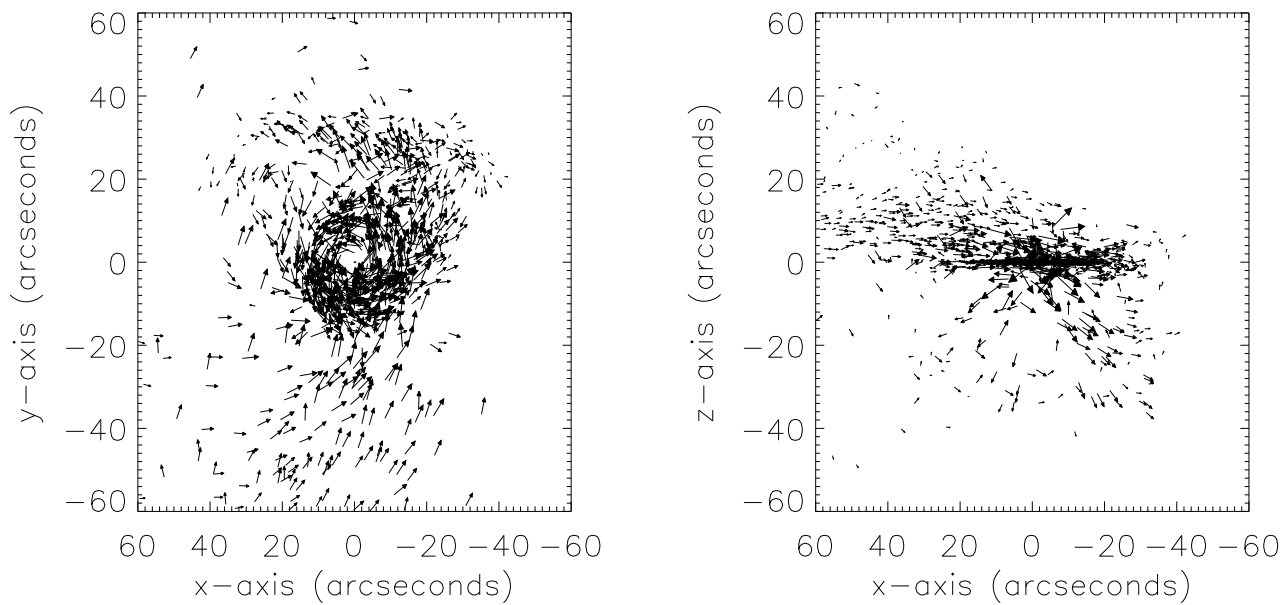}}
	\caption{Model velocity field. For clarity only 10\% of the
	vectors are shown.
	Left: model velocity field projected on the $x$-$y$-plane.
	The galaxy rotates counter--clockwise.
	The length of the arrows is proportional to the absolute
	value of the projected velocity. Right: model velocity field 
	projected on the $x$-$z$-plane.
	} \label{fig:model_velofield}
\end{figure*}

The projection of the cloud velocities on the $x$-$y$-plane and $x$-$z$-plane
are shown in Fig.~\ref{fig:model_velofield}.
The galaxy rotates counter-clockwise. The tenuous gas component at large 
distances from the galaxy falls back to the galaxy. The northern
shell consists of two counter-rotating streamers. The major streamer
comes from the west in the sense of the galactic rotation, the minor
one arrives from the east in a counter-rotating sense.
They correspond to the
two re-accretion arms which form after the stripping event when ram pressure
has already ceased. The first arm rotates in the sense of the galaxy rotation
and meets the disk in the south, the second arm is counter-rotating
and meets the galactic disk in the north. The northern shell is thus
expanding to the north. 

\subsection{Comparison with observations}

\begin{figure}
	\resizebox{\hsize}{!}{\includegraphics{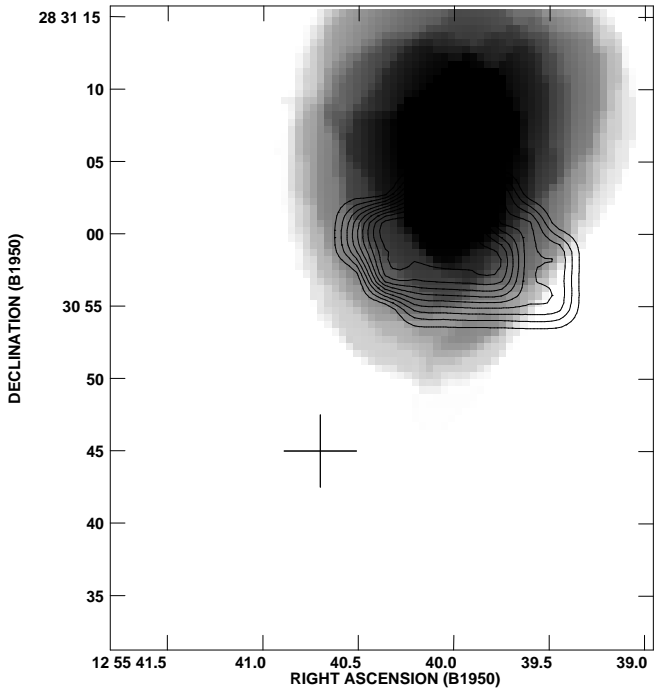}}
	\caption{Comparison between the simulation snapshot and the H{\sc i}
	observations of Bravo-Alfaro et al. (2000). Greyscale:
	model column density distribution. Contours: H{\sc i}
	emission distribution of Fig.~\ref{fig:co_10_HI}.
	} \label{fig:HI_model_obs}
\end{figure}

We have generated the observer's view of this gas distribution 
with PA=158$^{\rm o}$ and $i$=75$^{\rm o}$.  
For the comparison of the model column density distribution with the
H{\sc i} data we assume that the gas within a distance of 7 kpc
(15$''$) from the galaxy center is mainly molecular whereas the gas at
larger distances is mainly atomic. We can thus compare directly
both distributions (Fig.~\ref{fig:HI_model_obs}).
The low resolution H{\sc i} distribution of Bravo-Alfaro et al. (2000)
suggests that the bulk of the atomic gas is located more to the
north than the emission region shown in Fig.~\ref{fig:co_10_HI}.
The model H{\sc i} emission region is located in the right direction and at 
approximately the right distance from the galaxy center. Thus, we suggest that
the observed H{\sc i} emission regions corresponds to the inner part of 
the expanding northern shell of our simulation.

In order to compare the H$\alpha$ image
with the model, we again assume that star formation takes place
where the re-accreting material hits the disk, i.e. where we have a double
line profile. Each cloud complex in the model snapshot was checked 
for a velocity which leads to a collision with a neighboring
cloud complex. We thus project the velocity vectors  
(${\bf v_{1}}$, ${\bf v_{2}}$)
of each pair of clouds on their relative distance vector
${\bf r_{12}}$: $v_{1}^{\rm proj}={\bf v_{1}}\cdot
{\bf r_{12}}\ ;\ v_{2}^{\rm proj}={\bf v_{2}}\cdot {\bf r_{12}}\ .$
We apply a weighting factor to each cloud which is equal to
the number of possible collisions fulfilling the criterion:
$r_{12} \leq 1$~kpc, $v_{1}^{\rm proj}$ has the opposite sign of 
$v_{2}^{\rm proj}$.
All cloud complexes which will not collide are removed
from the cube. At distances $R \leq 7.5$ kpc, i.e. within
the ring observed in H$\alpha$, the mean free path of a cloud 
is smaller than outside the disk.
The criterion $r_{12} \leq 1$ kpc might thus not be valid at small distances.
Moreover, in the inner disk, star formation is more likely due to
density waves within the stellar and gaseous disk. Since this aspect is not 
included in the model we prefer to give the clouds at $R \leq 7.5$ kpc a
uniform weight (Vollmer et al. 2000). 
The model emission inside the galaxy disk is thus
density weighted, the emission outside the disk is collision weighted.

\begin{figure}
	\resizebox{\hsize}{!}{\includegraphics{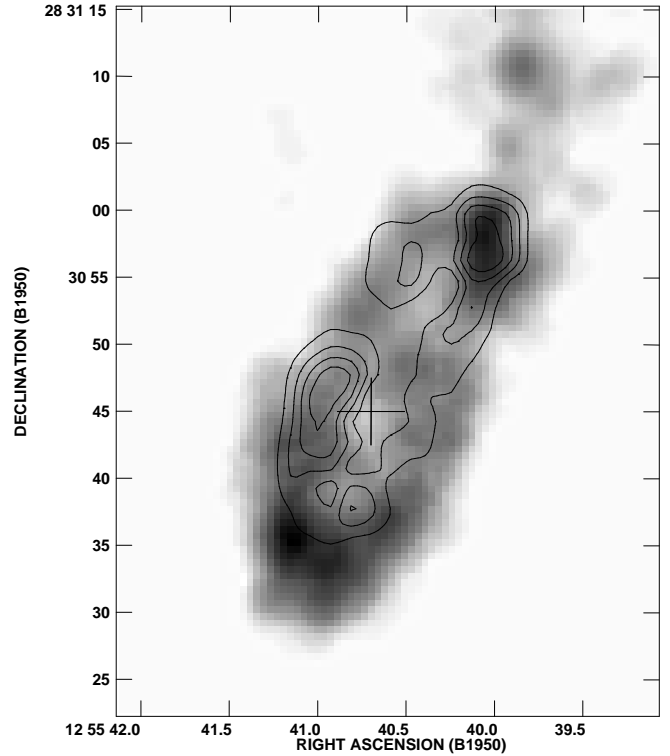}}
	\caption{H$\alpha$ emission distribution. Greyscale:
	Model. Contours: observations (Fig.~\ref{fig:co_10_Ha}).
	} \label{fig:model_counterrot}
\end{figure}

The result of this procedure together with the
H$\alpha$ emission distribution can be seen in 
Fig.~\ref{fig:model_counterrot}. 
The extent of the model and observed H$\alpha$ emission fits
remarkably well. The two brightest spots in the model are located 
in the north-west and in the south of the galaxy center.
Especially the north-western maximum coincides exactly
with the observed maximum of the H$\alpha$ emission.

\begin{figure}
	\resizebox{\hsize}{!}{\includegraphics{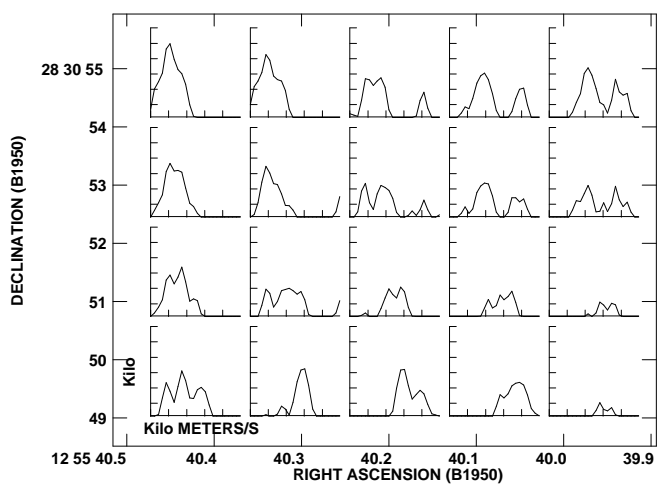}}
	\caption{Model H$\alpha$ spectra of the region in the west of the 
	galaxy. $x$-axis: velocity (6940 to 7468~km\,s${-1}$
	same as in Fig.~\ref{fig:counterrotation}); 
	$y$-axis: gas surface density in arbitrary units.
	} \label{fig:spectra_model}
\end{figure}

In order to check if we can reproduce the double line profile
in the north-eastern emission region as it has been observed by Amram et al.
(1992), we made spectra of the model cube in this
region (Fig.~\ref{fig:spectra_model}), which can be directly compared
with the H$\alpha$ spectra of Fig.~\ref{fig:counterrotation}.
As in the observed spectra, we find a double line profile in the model
cube.  The main difference between the model and observations 
lies in the different intensities of the lines.
The observed line of highest intensity is not due to regular rotation, 
whereas the model shows the contrary.  This difference is not significant,
however, because we have not included an explicit recipe for
star formation or radiative transfer in our model, making
it impossible to predict the outgoing ionizing UV field in the star 
forming region.

\subsection{Towards a coherent scenario}

Our model includes only 10\,000 particles but the calculation is 
very long for each run due to the small time steps, limiting
the exploration of parameter space.
Our aim is to show generic features which arise naturally 
from our simulations. These effects are more or less pronounced for
different $i$ and $p_{\rm max}$. The values chosen represent
a compromise as described in Sect. \ref{sec:model}. Nevertheless, there are 
typical features which appear only in a given range of $i$ and
$p_{\rm max}$, as the amount of re-accretion and the resulting H{\sc i}
deficiency. 

We are able to reproduce
\begin{itemize}
\item
the observed H{\sc i} distribution, i.e. that most of the H{\sc i} is found in the north 
of the galaxy,
\item
the north-western maximum of the observed H$\alpha$ emission distribution,
\item
the observed double H$\alpha$ line profile in the north-west of the galaxy,
\item
the enhanced 20 cm radio continuum emission region to the north of the galaxy center.
\end{itemize}

In our model the re-accreting gas hits the disk at the position of 
the observed CO emission region.  We expect that the molecular gas has condensed
from the colliding atomic gas, inducing the observed star formation.
The observed radio continuum peak near the CO/H{\sc i} emission region arises
naturally from an ISM-ISM collision in which the magnetic fields 
are compressed at the same time as the clouds.

\section{Alternative scenarios}

The scenario presented here, with the observational asymmetries of the gas
distributions and perturbations of the velocity fields, is not unique.
The distance to the Coma cluster is such that the resolution of the presented
observations is lower than that of the model. Therefore, we will discuss 
two alternative scenarios.

\subsection{Ram pressure stripping in action}

If ram pressure is stripping the gas now, we could imagine a scenario
where NGC~4848 approaches the cluster center. The neutral hydrogen 
and maybe a part of the molecular gas (Kritsuk 1983) are displaced to the north west 
of the galaxy by the ram pressure of the ICM. In this way NGC~4848
would resemble NGC~4419 in the Virgo cluster (Kenney et al. 1990), which also
shows an asymmetric distribution of the molecular gas. Nevertheless, there
are important differences. NGC~4419 is red $(B-V)_{\rm T}=1.0$ (Kenney et al. 1990), 
whereas NGC~4848 is a blue disk galaxy $(B-V)_{\rm T}=0.44$ (RC3). Moreover,
the asymmetry of the CO distribution is much more pronounced for
NGC~4419 (a factor of 2 in CO luminosity) than for NGC~4848 (a factor of 0.1). 
Thus, they might not be directly comparable.

For the scenario of active stripping, two main
questions remain: (i) The galaxy has lost about 2/3 of its atomic gas. Thus
ram pressure stripping has already been very efficient at larger
distances from the cluster center ($> 1$~Mpc) in order
to have removed more than $3\,10^{9}$~M$_{\odot}$. This gas must have been 
ionized, because we do not detect it in the 21~cm line. 
At these distances the ICM density is very low ($n_{\rm e} \sim 10^{-4}$~cm$^{-3}$).
If we assume an H{\sc i} column density of $\Sigma_{\rm HI}=10^{20}$~cm$^{-2}$, 
a rotation velocity $v_{\rm rot}=270$~km\,s$^{-1}$, and a galactic radius of 
$R$=10~kpc, we obtain from the energy density equation for 
face--on stripping (Gunn \& Gott 1972)
\begin{equation}
\Sigma_{\rm gas}v_{\rm rot}^{2}R^{-1}=m_{\rm p}n_{\rm e}v_{\rm gal}^{2}\ ,
\label{eq:gg72}
\end{equation}
where $m_{\rm p}$ is the proton mass, and we have a galaxy velocity of
$v_{\rm gal} \sim 1600$~km\,s$^{-1}$. The Keplerian velocity for
circular orbits at 1~Mpc from the cluster center, assuming an
enclosed total mass of $M_{\rm tot}=5\,10^{14}$~M$_{\odot}$ (Hughes 1989),
is $v_{\rm Kepler} \sim 1500$~km\,s$^{-1}$. If the galaxy was
stripped face--on, the scenario of active ram pressure stripping would
be consistent with the H{\sc i} 21~cm line observations.

NGC~4848 has a low radial velocity with respect to the cluster mean
($\Delta v \sim 350$~km\,s$^{-1}$). Thus, the galaxy's velocity vector
must have a large
component in the direction of the cluster center. Since it is highly
inclined, it is most probable that NGC~4848 is stripped nearly edge--on.
In this case, Eq.~\ref{eq:gg72} does not apply, i.e. ram pressure
must be much higher to strip the outer atomic hydrogen. According to
our model simulations (Fig.~\ref{fig:simulation1}), a 
displacement of the gas is only clearly detectable after the galaxy's
passage through the cluster center. Even if we underestimate the 
ram pressure efficiency, it is unlikely that a displacement
already takes place at $\Delta t \sim 400$~Myr before the galaxy's
closest approach to the cluster center.

\subsection{Tidal interaction with a dwarf galaxy \label{sec:tdwarf}}

Fig.~\ref{fig:Hband} shows an alignment of four nearly spherical emission
regions at an angle of $\sim 30^{\rm o}$ with respect to the galaxy's
position angle. This could be a tidally stretched dwarf galaxy crossing
the disk, in the same way the Sagittarius dwarf crosses the Galaxy
(see Edelsohn \& Elmegreen 1997 for numerical simulations).
In this case, one would expect a major emission region which
corresponds to the core of the dwarf galaxy. The two strongest emission
regions at the extremities of the alignment have approximately the same
brightness in the H band, weakening the interpretation of the alignment 
as a tidally stretched dwarf galaxy.

Even if the galaxy was gas rich, this can not explain the displacement
of $\sim 10^{9}$~M$_{\odot}$ of atomic hydrogen. Moreover, in an ISM--ISM
interaction one would expect strong H{\sc i} tails following the stellar
tails (see, e.g., Weliachew et al. 1978). An ISM--ISM collision
could explain the detached northern CO emission region together
with the enhanced radio emission, but it cannot account for
the higher H$\alpha$ line to continuum ratio in the west of the
H$\alpha$ ring, either for the large spatial extent of the region
where H$\alpha$ double line profiles were observed, or for the 
offset between the alignment and the western H$\alpha$ maximum.

As ram-pressure stripping is necessary in the two 
possible scenarios and naturally accounts for the observations according to
our simulations, it is currently the preferable explanation.  Further 
observations (e.g., with the HST) 
may allow us to determine whether a small galaxy is also being tidally torn apart.

\section{Conclusions}

We have shown $^{12}$CO(1--0) Plateau de Bure observations of the highly
H{\sc i} deficient Coma galaxy NGC~4848. 
A detached CO emission region in the north of the galaxy
center was detected. This region is located near an H$\alpha$ emission
maximum where an H$\alpha$ double line profile was observed earlier.
Furthermore, an enhancement in the 20~cm radio continuum coincides
with the detached northern CO emission region.

We are able to reproduce naturally the main characteristics of the
emission distributions and velocity fields of the
multiple wavelength observations with the help of a numerical simulation
including ram pressure stripping. In this scenario, the galaxy passed
near the cluster center $\sim$400~Myr ago.  
NGC~4848 is emerging from the cluster center with a radial velocity of 
$\sim$-400 km\,s$^{-1}$ and is located in front of the cluster center.
We propose that it was an Sc galaxy which entered the cluster for
the first time $\sim$1~Gyr ago. During its recent passage through the 
cluster core most of its atomic gas was removed by ram pressure
stripping. Its molecular gas has not been stripped because it is
located deep in the gravitational potential well of the galaxy 
and because of its very high column density. 
The atomic gas which was not accelerated to a velocity above the 
escape velocity now falls back onto the galaxy. The compression of the 
neutral gas due to accretion leads to a phase transition from atomic to
molecular gas and to star formation activity.  We expect the star formation 
rate to decrease within the next Gyr due to the lack of a  gas reservoir 
in the outer disk.

An alternative scenario, where ram pressure stripping is active and 
a tidally stretched dwarf galaxy crosses the disk at the same time, 
is also consistent with the available data but less probable
on the basis of our numerical simulations.

\begin{acknowledgements}
We would like to thank the IRAM staff for their kind support during the
data reduction phase in Grenoble and P. Amram and M. Marcelin for making
the H$\alpha$ data cube available to us. 
BV was supported by a TMR Programme of the European Community
(Marie Curie Research Training Grant). 
We would also like to thank J. Kenney for helping us to improve this
article significantly.
\end{acknowledgements}

\end{document}